\documentclass[a4paper, 12pt]{article}

\usepackage[latin1]{inputenc}
\usepackage[english]{babel}

\usepackage{graphicx}
\usepackage{amsmath}
\usepackage{latexsym}
\usepackage{amssymb} 
\usepackage{epsfig}
\usepackage{cancel}
\usepackage{exscale}
\usepackage{color}

\newcommand{\D}{\displaystyle}
\newcommand{\T}{\textstyle}

\newcommand{\SSi}{\scriptscriptstyle}
\newcommand{\p}{\varphi}

\begin{document}

\begin{titlepage}
$\phantom{absatz}$
\hspace{9.5cm}
Bonn-TH-09-02
\vspace{1.5cm}
\begin{center}
{\Large \bf Proof of the Lagrangean formalism of Hermitean 1-matrix models to
all orders}   
\vspace{2.0cm}

{ \large Alexander Klitz}
\\ 
\vspace{1.5cm}
Physikalisches Institut, Universit\"at Bonn,\\
Nu{\ss}allee 12, 53115 Bonn, Germany\\ 
\vspace{1cm}

\begin{abstract}
 We show that the correlation functions and the free energy of the formal Hermitean
 1-matrix model can be described by the recently proposed Lagrangean
 formalism to all orders. In addition, the loop equation of this formalism is
 stated and solved up to the fifth order in $ 1 / N^2 $ explicitly.
\end{abstract}

\end{center}
\end{titlepage}

\newpage
\section{Introduction}
\label{section: Introduction}
For matrix models with fixed filling fractions, substantial progress was made
in 2004 by Eynard \cite{E-04}. By utilising loop equations of higher degree, it
was possible to write a recursion formula for the correlation functions. Since
this formula contains one residue, a single term of one specific correlation
function is given by a system of nested residues. A diagrammatic
representation for these terms was developed in \cite{E-04} and subsequent
work \cite{CE-06}, \cite{EO-07}. These diagrams were initially claimed to be
Feynman diagrams \cite{CE-06}. Later this assertion was revoked \cite{EO-07},
\cite{EO-08}. In the context of topological string theory and the topological
B model, Dijkgraaf and Vafa \cite{DV-07} found an action for a chiral theory
which leads to the recursion equations and diagrammatics of \cite{EO-07}.\\ \\
In \cite{FGK-08}, Hermitean 1-matrix models were described by an effective
field theory. For the planar approximation of the $ 1/N $-expansion, a proof
was given and several examples of the first and second order corrections
with nontrivial symmetry factors were checked. The present work
compares in general the number of Wick contractions in the field theory with
the solution of the loop equations and the action of the loop operator in the
matrix model. The equality of both approaches to all orders is a proof of the Lagrangean formulation. \\ 

The free energy was obtained in \cite{FGK-08} by using some informations about the so-called H-operator, which was introduced by Eynard, Chekhov and
Orantin \cite{CEO-06} for the 2-matrix model and then in the 1-matrix model
\cite{CE-06}. We describe a new method for the calculation of the free energy, which does not rely on the operator $ H $. This new approach is an independent confirmation of the previously
obtained result. With both approaches, using the
operator H or not using it, one can modify the theorem stated in \cite{FGK-08}
to incorporate the free energy directly into the theorem. In this case, one has to introduce $
\mathcal{L}_0^{(h)} $ for $ h \geq 2 $. With an operator defined for tree
graphs in \cite{G-06}, one can give an explicit formula for $ \mathcal{L}_k^{(h)}
$. This also simplifies the loop formula---only one instead of $ ( 3 h - 1 ) $
couplings has to be calculated. In this loop equation, only simple projection
operators and differential
operators appear, which makes the calculation of higher order contributions simple.\\

Apart from directly using the Lagrangean formalism for the calculation of
correlators for matrix models, one can identify further applications.\\
Critical phenomena related to matrix models (\cite{BZ-92}, \cite{HINS-93},
\cite{HINS-95}, \cite{HINS-97}) can be explored by the
renormalization group of the Lagrangean formalism. Another application deals
with the enumeration of discrete surfaces (also called maps) (\cite{T-63},
\cite{BIPZ-78}, \cite{EO-08}). A third
possibility is the calculation of Weil-Petersson volumes with the effective
Lagrange density \cite{EO-07-A}. This scheme can also be applied to topological string theory.

The plan of the paper is as follows: In section \ref{section: Definitions and
  notations}, the correlation functions
and other relevant quantities are defined. A list of the equations necessary
for the proof which were already derived in \cite{FGK-08} concludes this
part. In section \ref{section: Lagrangean formalism}, theorem $ 1 $ and the equivalent theorem $ 1a $ are formulated. Both directly include the free energy. A new, independent method of obtaining the free energy is
described and the loop equation of the Lagrangean formalism is stated. In sections \ref{section: Induction on h} and \ref{section: Induction on k}, we prove theorem $ 1 $ and theorem $ 1a $. 

\section{Definitions and notations}
\label{section: Definitions and notations}
The free energy $ \mathcal{F} $ and the partition function $ \mathcal{Z}_N (t)
$ of the formal Hermitean matrix model are defined by an integral over
Hermitean matrices
\begin{eqnarray}
\label{eq: partition sum}
 \mathcal{Z}_N (t) = \int  e^{-N \; {\rm tr} \; V(M) } \; dM = e^{ \mathcal{F}
   }\;,
\end{eqnarray}
\begin{displaymath}
V ( M ) = \sum_{n \ge 1 } t_n M^n .
\end{displaymath}

The formal expansion of the free energy is given by

\begin{equation} 
 \mathcal{F}= \sum_{h=0}^\infty N^{2-2h} \; \mathcal{F}^{(h)}.  
\end{equation}

To give two equivalent definitions of correlation functions, we need the resolvent

\begin{displaymath}  
W(z) = \frac{1}{N} tr \frac{1}{z-M} \;,  
\end{displaymath}

the loop operator \cite{A-96}

\begin{equation} 
\label{eq: loop operator}
 \frac{ \partial }{ \partial V (z) } = - \sum_{j=1}^\infty \frac{1}{z^{j+1}} 
\frac{\partial}{\partial t_j} 
\end{equation} 

and a notion of an average of $ f $

\begin{displaymath}
 \Big<  f(M)  \Big>  = \frac{1}{\mathcal{Z}_N (t)} \int f(M) \; e^{-N \; {\rm tr} \; V(M) }
 \; dM  .
\end{displaymath}

Then the $ n $-point function reads
\begin{displaymath}   
W_n (z_1, \ldots ,z_n) = N^{2n-2} \Big< W(z_1) \ldots W(z_n) \Big>_{\rm conn} = \frac{1}{N^2}
\frac{ \partial }{\partial V (z_1) } \cdots \frac{ \partial }{\partial V (z_n)
  } \mathcal{F} 
\end{displaymath}
for $ n \in \mathbb{N} $. The $ 1 / N $-expansion of the free energy translates into a $ 1/N
$-expansion of the correlation functions

\begin{displaymath}
 W_n (z_1, \ldots, z_n) = \sum_{h=0}^\infty N^{-2h}  W_n^{(h)} (z_1, \ldots, z_n)
\end{displaymath} 

for $ n \in \mathbb{N}_0 $. (This implies $ W_0^{(h)} = \mathcal{F}^{(h)} $ and $
W_0 = \mathcal{F} / N^2 $.)

The spectral curve

\begin{equation}
\label{eq: y}
 y (x) = V^\prime (x) - 2 W_1^{(0)} (x) 
\end{equation}

is an essential concept underlying the matrix model. Vertex factors of the Lagrangean
formalism consist exclusively of integrals in $ y $.
The polynomial \cite{A-96}

\begin{equation} 
 M(x) = \oint\limits_\mathcal{C_\infty} \frac{d \omega}{2 \pi i} \frac{1
 }{\omega - x } 
\frac{V^\prime (\omega) }{\sqrt{\prod_{i=1}^{2s} (\omega - a_i ) }}   
\end{equation}

divides the spectral curve $ y(x) = M(x) \tilde{y} (x) $ into the reduced spectral curve $ \tilde{y} $:

\begin{equation}
\label{eq: reduced y}
\tilde{y}^2 (x) = \prod_{i=1}^{2s} (x - a_i).
\end{equation}

$ \mathcal{C}_\infty $ denotes a contour encircling counterclockwise all $ s $
cuts
$ [ a_1, a_2 ], \ldots ,[ a_{2s-1} , a_{2s} ] $; $ A_l $ denotes a contour
encircling counterclockwise only the $ l $-th cut $ [ a_{2l-1} , a_{2l} ]
$. The following $ s-1 $ equations for the filling fractions $ \varepsilon_l
$, which are given parameters of the theory, fix the $ s-1 $ residual degrees of freedom:
 
\begin{displaymath}
 \varepsilon_l = \frac{1}{2} \oint\limits_{A_l} \frac{d x }{2 \pi i} y(
 x)  \mbox{ for } l = 1,\ldots, s-1,
\end{displaymath}
\begin{displaymath}
\sum_{l=1}^s \varepsilon_l = 1 .
\end{displaymath}

A second essential ingredient to build the correlators of the matrix model is the Bergmann kernel
\begin{equation}
\label{eq: relation between B and dS}
 B(p,q) = \frac{1}{2} \frac{\partial}{\partial q} \left( \frac{ dp }{ p-q }
  + dS (p,q) \right) dq . 
\end{equation}

The differential $ dS (x,x^\prime) $ is given by

\begin{equation}
\label{eq: explicit expression for dS }
 dS(x, x^\prime) = \frac{\sqrt{\sigma(x^\prime)}}{\sqrt{\sigma(x)}}  \left(    
  \frac{1}{x-x^\prime} - \sum_{j=1}^{s-1} C_j (x^\prime) L_j (x) \right) dx 
\end{equation} 
\begin{displaymath} 
 \mbox{with\ \ } C_j (x^\prime) = \frac{1}{2 \pi i} \oint\limits_{A_j}  
\frac{dx}{\sqrt{ \sigma(x) }} \frac{1}{x-x^\prime} \mbox{ and } \sigma(x) =
\prod_{i=1}^{2s} (x-a_i) .  
\end{displaymath} 

The $ s-1 $ polynomials $ L_j (x) $ of degree $ s-2 $ are uniquely determined
by the conditions

\begin{displaymath}
 \oint\limits_{A_l}  \frac{L_j(x)}{\sqrt{ \sigma(x) }} dx  = 2 \pi i \delta_{jl}
 \mbox{\ for 
  \ } l,j = { 1,\ldots,  s-1}.  
\end{displaymath}

We denote
\begin{displaymath}
 \widetilde{B} (p,q) = \frac{ B(p,q) }{ dp \; dq } - \frac{1}{2}
\frac{1}{(p-q)^2} .  
\end{displaymath}

The propagators for external lines $ B_i^f (p) $ ($ f = 0,1,2,\ldots $), the propagators for internal
lines $ B_{i,j}^{f,g} $ ($ f,g=0,1,2,\ldots $) and the constituents of the
vertex factors $ y_{f,i} $ ($ f=1,2,3,\ldots$) are defined as follows:

\begin{eqnarray*}
 B_i^f (p) & = & \setlength{\unitlength}{0.250cm}
\begin{picture}(0,4)
\put(1.4,3.3){\line(1,0){3.9}}
\put(1.4,3.3){\line(0,-1){1.0}}
\put(5.3,3.3){\line(0,-1){1}}
\end{picture}
\p ( p ) ( \partial^{f} \p )_{ i } 
= 2 \oint\limits_{a_i} \frac{dz}{ 2 \pi i } \frac{
   \widetilde{B} ( p, z ) }{ ( z - a_i )^{f+1/2} } \\
B_{i,j}^{f,g} & = & \setlength{\unitlength}{0.250cm}
\begin{picture}(0,4)
\put(2.0,3.3){\line(1,0){4.7}}
\put(2.0,3.3){\line(0,-1){1.0}}
\put(6.7,3.3){\line(0,-1){1}}
\end{picture}
( \partial^f \p)_i ( \partial^{g} \p )_{ j } =
4 \oint\limits_{a_i} \frac{dz}{ 2 \pi i } \oint\limits_{a_j} \frac{
  dz^\prime }{ 2 \pi i } \frac{ \widetilde{B} (z,z^\prime) }{(z-a_i)^{f+1/2}
  (z-a_j)^{g+1/2} } \\
y_{f,i} & = & \oint\limits_{a_i} \frac{dz}{ 2 \pi i} \frac{ y(z) }{ (z -
  a_i)^{f+1/2} } .
\end{eqnarray*}

As was shown in \cite{FGK-08}, the lowest correlation functions, in particular
$ W_3^{(0)} $ and $ W_1^{(1)} $, are composed of the aforementioned
propagators and vertex factors. When the loop operator is applied to such
expressions it can be reformulated as

\begin{displaymath}
\frac{ \partial }{ \partial V (q) } = \sum_{k,i,f} \frac{ \delta B_i^f (p_k)
}{ \delta V (q) } \frac{ \partial }{ \partial B_i^f (p_k) }  
+ \sum_{f,g,i,j} \frac{ \delta B_{i,j}^{f,g} }{ \delta V (q) } \frac{ \partial
}{ \partial B_{i,j}^{f,g} }
+  \sum_{f,i} \frac{ \delta y_{f,i} }{ \delta V (q) } \frac{ \partial
}{ \partial y_{f,i} } .
\end{displaymath}

In this way the loop operator can be subdivided into the following seven parts \cite{FGK-08}:

\begin{eqnarray}
\label{eq: y_fi variation}
\frac{ \delta y_{f,i}}{\delta V (q) } &=& \big( \Delta_1 (q) + \Delta_2 (q) 
\big) (y_{f,i}) 
\\
  &&\big( \Delta_1 (q) \big) (y_{f,i})= (2 f + 1) \frac{ y_{f+1,i}} {y_{1,i}}  
B_i^0 (q) \ \ \ \ \ \ \ \ \ \ \ \ \ \ \ \ \  {\rm \phantom{(\ref{eq: y_fi variation}a)}} 
\nonumber\\
&&\big( \Delta_2 (q) \big) (y_{f,i}) = - B_i^f (q) \ \ \ \ \ \ \ \ \ \ \ \ \ \
\ \ \ \ \ \ \ \ \ \ \ \ \ \ \ \ \ \ \ {\rm \phantom{(\ref{eq:
    y_fi variation}b)}} 
\nonumber
\end{eqnarray}
\begin{eqnarray}
\label{eq: B_i^f variation} 
\frac{ \delta B_i^f (p) }{\delta V (q) } &=& \big( \Delta_3 (q)  +  \Delta_4
(q) \big) (B_i^f (p))
\\
&&\big( \Delta_3 (q) \big) (B_i^f (p))= (2 f + 1) \frac{ B_i^{f+1} (p) B_i^0
  (q) }{ y_{1,i} } \ \ \ 
\ \ \ \ \ \ \ {\rm \phantom{(\ref{eq: B_i^f variation} 
a)}}
\nonumber\\
&&\big( \Delta_4 (q) \big (B_i^f (p)) =  \sum_{j=1}^{2s} \frac{ B_{i,j}^{f,0}
  B_j^0 (p) B_j^0 (q) }{ y_{1,j} } \ \ \ \ \ \ \ \ \ \ \ \ \ \ \: {\rm \phantom{(\ref{eq: B_i^f variation}
b)}}
\nonumber\\
\label{eq: B_ij^fg variation} 
\frac{ \delta B_{i,j}^{f,g} }{ \delta V (q) }  &=& \big( \Delta_5 (q) + \Delta_6 (q)
+ \Delta_7 (q) \big) ( B_{i,j}^{f,g})
\\
&&\big( \Delta_5 (q) \big) ( B_{i,j}^{f,g}) = ( 2 f + 1 ) \frac{ B_{ i,j 
  }^{ f+1,g } B_i^0 (q) }{ y_{1,i} } \ \ \ \ \ \ \ \ \ \ \ \ \ \ \, {\rm \phantom{(\ref{eq: B_ij^fg variation}a)}}
\nonumber\\
&&\big( \Delta_6 (q) \big) ( B_{i,j}^{f,g}) = ( 2 g + 1 ) \frac{ B_{ i,j }^{f,g+1} 
  B_j^0 (q) }{ y_{1,j} }  \ \ \ \ \ \ \ \ \ \ \ \ \ \ \ \! \! \; {\rm \phantom{(\ref{eq: B_ij^fg variation}b)}}
\nonumber\\
&&\big( \Delta_7 (q) \big) ( B_{i,j}^{f,g}) = \sum_{k=1}^{2s} \frac{ B_{i,k}^{ f,0 } B_{ j,k }^{ g,0 }
  B_k^0 
  (q) }{ y_{1,k}} .\ \ \ \ \ \ \ \ \ \ \ \ \ \ \ \ \ \! \: {\rm \phantom{(\ref{eq: B_ij^fg variation}c)}}
\nonumber
\end{eqnarray}

An application of $ \Delta_j (q) $, $ j=1,\ldots,7 $ to any expression not
noted above gives zero.
Utilising the loop equations of the matrix model, one can find \cite{E-04} a
recursion equation for the correlation functions

\begin{equation}
\label{eq: loop eq. to order
h}
 W_1^{(h)} (p) = \sum_{i=1}^{2s}  \underset{ x \rightarrow a_i }{\rm Res}
 \frac{ dS(p,x) }{ dp } \frac{1}{y(x)} 
\left( \sum_{m=1}^{h-1} W_1^{(h-m)} (x)  W_1^{(m)} (x) + W_2^{(h-1)} (x,x)
\right)  . 
\end{equation}

This formula can be evaluated by using the equation \cite{FGK-08}

\begin{eqnarray}
&\underset{ x \rightarrow a_i }{\rm Res }& \frac{ dS (p,x) }{dp\; y(x) } B_j^f 
(x) B_k^g (x)  
\nonumber\\
&=&  \frac{1}{2} B_{j,i}^{f,0} \frac{ B_i^0 (p) }{ y_{1,i} } 
B_{i,k}^{0,g} 
\nonumber\\
&+& \delta_{k,i} (g+\frac{1}{2}) \sum_{ r + m + t = g + 1 } \frac{ 1 }{ 2 r + 1 } 
B_i^r (p)  B_{j,i}^{f,m} Z_{t,i}
\nonumber\\ 
&+& \delta_{j,i} (f+\frac{1}{2}) \sum_{ r + m + t = f + 1 } \frac{ 1 }{ 2 r + 1 }
B_i^r (p)  B_{k,i}^{g,m} Z_{t,i}
\nonumber\\
\label{eq: integral contour twist result}
&+& \delta_{k,i} \delta_{j,i} \frac{1}{2} ( 2 g+1 ) (2 f + 1) \sum_{ r + l = f +
  g +2 } \frac{1 }{2r+1} B_i^r (p) Z_{l,i} 
\end{eqnarray}

with

\begin{eqnarray*}
Z_{n} = \sum_{ k_1 + \ldots + k_n = k \atop \sum_j j k_j = n } \frac{ ( -
  )^{k} k! }{ k_1! \cdots k_n ! } \frac{ 1 }{ (y_{1})^{k+1} } \prod_{l=1}^n (
y_{1+l} )^{k_l}  = \sum_{k=0}^n \frac { Z_{n}^{[k]} }{ (
  y_{1} )^{k + 1 } } .
\end{eqnarray*}

$ Z_{n,i} $ is obtained by adding to each $ y_f $ in $ Z_n $ the index $ i $
giving $ y_{f,i} $. This corresponds to specifying the general coordinate $ x
$ as $ x=a_i $.

\section{Lagrangean formalism}
\label{section: Lagrangean formalism}
The different terms of the Lagrange density with the topological index $ h $ read
\begin{eqnarray*}  
\mathcal{L}_k^{(h)}  (a_i) = \sum_{ \alpha \in M_k^{(h)} }
\lambda^{(h)}_{\alpha,i} \; \frac{ \p_i^{\alpha_0} ( \partial^1
  \p)_i^{\alpha_1} \cdots (\partial^{k+3 h - 3}   \p)_i^{\alpha_{k+3
      h -3}} }{ \alpha_0 !\ \  \alpha_1 ! \ \ \ \ \cdots \ \ \ \ \ \
  \alpha_{k+3 h -3} ! \ \ \ \ \;} = \sum_{ \alpha \in M_k^{(h)} }
\lambda^{(h)}_{\alpha,i} \; \frac{ \p_i^{\alpha} }{ \alpha ! } .
\end{eqnarray*}
The set $ M_k^{(h)} $ consists of all multi-indices $ \alpha = ( \alpha_0 , \ldots ,
\alpha_{k+3h-3} ) \in ( \mathbb{N}_0 )^{k+3h-2} $ which fulfill the
conditions $ \sum_{j=0}^{k+3h-3} j \alpha_j \leq k+3h-3 $ and $
\sum_{j=0}^{k+3h-3} \alpha_j = k $.
\begin{equation} \lambda_{ \alpha  }^{(h)} = \left( \, \underset{ f = 1 }{
      \overset{ k + 3 h - 3 } {\mathnormal{ \prod }} } \left( - \frac{ \partial }{ \partial y_{f}
    } \right)^{\alpha_f} \right) \left( \sum_{f=1}^\infty ( 2f +1 ) \frac{
    y_{f+1} }{ y_1 } \frac{ \partial }{ \partial y_{f} } \right)^{ \alpha_0 - 3 \, \delta_{h,0}}
\; \lambda^{(h)} \equiv D_\alpha \; \lambda^{(h)}
\label{eq: lambda}
\end{equation}
The operator notation $ D_\alpha $ was first used in \cite{G-06}, in the special case of
tree diagrams. The two parts are the remnants of $ \Delta_1 $ and $
\Delta_2 $ freed from their propagators. This operator, as was discussed in
(\cite{FGK-08}, page 12), is sufficient to describe the action of the loop
operator on the level of the Lagrange density for the 1-vertex diagrams with
exclusively external legs. How to obtain the following list of  coupling constants $
\lambda^{(h)} $ is described in the next subsection on the loop equation.
\begin{eqnarray*}
\lambda^{(0)} &=& \frac{1}{y_1} \\
\lambda^{(1)} &=& - \frac{1}{24} \; {\rm log} \; y_1 \\
\lambda^{(2)} &=& -\frac{21}{160} \; \frac{ y_2^3 }{ y_1^5 } +
\frac{29}{128} \; \frac{y_2 y_3}{y_1^4} - \frac{35}{384} \; \frac{y_4}{y_1^3}
\\
&\vdots&
\end{eqnarray*}
$ \lambda_{ \alpha , i }^{(h)} $ is obtained by adding to each $ y_f $ ($ f =1
, 2, 3 ,\ldots $) in $ \lambda_{ \alpha }^{(h)} $ the index $ i $ giving $
y_{f,i} $.
\begin{eqnarray*}
\mathcal{L}^{(0)} &=& \phantom{ \mathcal{L}_0^{(0)}  + 
  \mathcal{L}_1^{(0)}  + \mathcal{L}_2^{(0)} +} \ \ \! \; \mathcal{L}_3^{(0)} + \mathcal{L}_4^{(0)} + \ldots
\\
\mathcal{L}^{(1)} &=& \phantom{ \mathcal{L}_0^{(1)} } \phantom{ + } \ \ \! \; \mathcal{L}_1^{(1)} +
\mathcal{L}_2^{(1)} + \mathcal{L}_3^{(1)} + \mathcal{L}_4^{(1)} + \ldots
\\
\mathcal{L}^{(h)} &=& \mathcal{L}_0^{(h)} + \mathcal{L}_1^{(h)} +
\mathcal{L}_2^{(h)} + \mathcal{L}_3^{(h)} + \mathcal{L}_4^{(h)} + \ldots
\mbox{ for } h \geq 2
\end{eqnarray*}

We would like to exclude four special cases from the correlators and define
\begin{displaymath}
\overline{W}_k = W_k - \delta_{k,0} ( \mathcal{F}^{(0)}  + \frac{ 1}{N^2}
\mathcal{F}^{(1)} ) - \delta_{k,1} W_1^{(0)} - \delta_{k,2} W_2^{(0)}.
\end{displaymath}

For the two special free energies, we refer to \cite{EO-07} and \cite{C-04}. The two special correlators are described directly by the spectral curve and the Bergmann kernel \cite{E-04}. The expression $ \langle 0 | \ldots | 0 \rangle_{l \ \rm loops} $ denotes the sum of all $
l $ loop diagrams of $ \langle 0 | \ldots | 0 \rangle_{conn} $.\\ \\

\underline{\bf Theorem $ 1 $} \\ 

With the Lagrange density $ \mathcal{L} = \mathcal{L}^{(0)} + \frac{ 1 }{ N^2 }
\mathcal{L}^{(1)} + \frac{ 1 }{ N^4 } \mathcal{L}^{(2)} + \ldots $ the
Hermitean 1-matrix model correlation functions and the free energy ($
\mathcal{F}^{(h)} \equiv W_0^{(h)} $) are given by
\begin{eqnarray*}
\overline{W}_k (p_1, \ldots, p_k ) =\sum_{l=0}^\infty \frac{ 1 }{ N^{ 2l } }
\left< 0 \left| \; \p(p_1) \cdots \p(p_k) e^{ \sum_{i=1}^{2s} \mathcal{L}
(a_i) } \right| 0 \right>_{l \ {\rm loops}}.
\end{eqnarray*}\\ \\

This theorem implies that the correlation function $ W_k^{(h)} $ consists of all connected diagrams with $ k $ external legs where the sum of topological indices $ h_j $ of the vertices plus the number of loops $ l $ in the diagram is equal to $ h $: $ l + \sum_{j} h_j = h $. \\
To proof theorem $ 1 $, the following equivalent formulation as theorem $ 1a $ is more
appropriate. The structure $ \mu $ of a vertex is defined by $ \mu = ( h ,
\alpha ) $, where $ h $ is the topological index of the
vertex and $ \alpha = ( \alpha_0 ,
\alpha_1 , \alpha_2 , \ldots ) $ determines with $ \alpha_j $ the number of emanating lines with a $ j $-th
derivative.
In a given diagram $ D $, with $ n $ vertices, the vertex $ j $ has a vertex
structure  $ \mu_j =  ( h_j , \alpha ( \mu_j ) ) $. The internal line $ x $
connects vertex $ i(x) $ with $ k(x) $ derivatives to vertex $ j(x) $ with $
l(x) $ derivatives. The external line $ v $ connects $ p_v $ to vertex $ d(v)
$ with $ m(v) $ derivatives. Then we define
\begin{displaymath} 
S_D := \sum_{ i_1, \ldots , i_n = 1 }^{2s} \Big( \prod_{ \rm external \atop
  line \ \it v = 1
}^{\SSi k} 
B_{ i_{d(v)}}^{ m(v) } ( p_v )  \Big)   \Big( \prod_{ \rm internal
  \atop line\ \it x }  
B_{ i_{ i(x)} , i_{j(x)} }^{k(x) , l(x)}
 \Big) \Big( \prod_{ \rm vertex \atop \it j=1 }^{ \SSi n} \lambda^{(h_j)}_{ \alpha (
  \mu_j ), i_j } \Big)  .
\end{displaymath}

$ \Pi_D $ is the inverse symmetry factor of diagram $ D $ (\cite{PS-95}, \cite{S-93}).\\
The summation $ \sum_{ \rm diagrams \ \it D} $ consists of all diagrams
with $ k $ external lines fulfilling $ l + \sum_{j=1}^n h_j
= h $, where $ l $ is the number of loops. \\ \\

\underline{\bf Theorem $ 1a $} \\ \\
For $ (k,h) \in \mathbb{N}_0 \times \mathbb{N}_0 \setminus \{ (0,0) , (1,0),(2,0),(0,1)
\} $, the Hermitean 1-matrix model correlation functions of order $ h $ and the free energy of
order $ h $ are given by
\begin{displaymath}
W_k^{(h)} ( p_1, \ldots, p_k ) = \sum_{\rm diagrams \ \it D} \Pi_D S_D.
\end{displaymath} \\ \\

In appendix B, all diagrams of $ \mathcal{F}^{(2)} $ and $ W_2^{(1)} $ emerging from the above theorem $ 1a $ are depicted.\\ 

{\large \bf Loop equation} \\ 

The form of $ \lambda^{(0)} $ is evident from inspection of $ W_3^{(0)} $, whereas $
\lambda^{(1)} $ has to be guessed. The guess is correct if $ D_\alpha
\lambda^{(1)} $ for $ \alpha = (1,0) $ and $ \alpha = (0,1) $ correctly
describes two of the three terms of $ W_1^{(1)} $ calculated in
\cite{FGK-08}. \\The $ \lambda^{(h)} = \lambda^{(h)}_{(0)} = \lambda^{(h)}_{(0)} ( y_1, \ldots , y_{
  3 h - 2 } ) $ for $ h \ge 2 $ are determined in a recursive process using loop
  equations. The loop equation (eq. (\ref{eq: loop eq. to order
h})) then becomes, in this Lagrangean formalism, an equation
for the coupling constants:
\begin{eqnarray} 
\hspace{-0.7cm}\lambda_{ \alpha^\prime }^{(h)} =
\sum_{m=1}^{h-1} \sum_{ \alpha \in M_1^{(h-m)} } \sum_{ \beta \in M_1^{(m)} } 
 \frac{ 2 A_\alpha A_\beta }{ 2 k^\prime +1 } 
Z_{n(\alpha,\beta,m) - k^\prime + 1  }
 \lambda_{\alpha }^{(h-m)}   \lambda_{ \beta  }^{(m)} 
\nonumber\\
  + \sum_{ \alpha \in M_2^{(h-1)} } \left( 2 - \sum_{ f=0 }^{ 3h-4 } \delta_{
      \alpha_f , 2 } \right) 
\frac{ 2 A_\alpha }{ 2 k^\prime +1 } Z_{ n(\alpha) - k^\prime + 1}  \lambda_{ \alpha  }^{(h-1)}
\end{eqnarray}

where $ A_\alpha = \prod_{f=0}^\infty ( f + 1/2 )^{ \alpha_f } $, the first index $ n(\alpha) = 1 + \sum_{j=0}^{3h-4}
    j \alpha_j  $, \\ the second index $ n(\alpha,\beta,m)= 1 + \sum_{j=0}^{3(h-m)-2} j \alpha_j
    + \sum_{j=0}^{3m-2} j \beta_j    $

and $  \alpha^\prime_j = \delta_{j,k^\prime} $ with $ k^\prime = 0 , \ldots , 3h-2 $. 
Since there are no $ y_1 $ independent parts of $ \lambda_{ \alpha^\prime
}^{(h)} $, one can deduce by induction over $ h \geq 2 $:
\\ \phantom{A}\ \ \ \ \ \ \ \ \ \ \ \ $ \lambda^{(h)} = $
antiderivative of $ ( - \lambda_{ ( 0 1 0 \cdots 0 )  }^{(h)} ) $ with respect
to $ y_{1} $. \\Since the integration is elementary, one also can
eliminate the integration in the equation.

The described procedure is a new, independent method of obtaining $ \widehat{
  \mathcal{F} }^{(h)} = \sum_{i=1}^{2s} \lambda_{ ( 0  ),i }^{(h)} $
and hence $ \mathcal{F}^{(h)} $, which does not rely on the operator $ H $ of reference \cite{CE-06}.\\

We obtain, after integration for $ h \ge 2 $,
\begin{eqnarray}
\hspace{-0.7cm}\lambda^{(h)} =
\sum_{m=1}^{h-1} \sum_{ \alpha \in M_1^{(h-m)}    }
\sum_{ \beta \in M_1^{(m)} }
 \sum_{ k, r, r^\prime  = 0 }^\infty  
\frac{ 2 A_\alpha A_\beta }{ 3 ( k + r + r^\prime  )  }
\frac{ Z_{n(\alpha,\beta,m)}^{[ k ]} }{ ( y_1 )^{k } }   ( P_r
D_\alpha \lambda^{(h-m)} ) ( P_{ r^\prime }
D_\beta \lambda^{(m)} ) 
 \nonumber\\
\label{eq: Lagrangean loop eq}
 + \sum_{ \alpha \in M_2^{(h-1)} }      \sum_{k,r=0}^\infty \left( 2 - \sum_{ f=0 }^{ 3h-4 } \delta_{
      \alpha_f , 2 } \right) \frac{  2 A_\alpha }{ 3 (  k + r ) }  \frac{ Z_{
    n(\alpha) }^{ [ k ] } }{ ( y_1 )^{ k } }   ( P_r
D_\alpha \lambda^{(h-1)} ) \ \ \ \ \ \  
\end{eqnarray}
where $ P_r $ is a projection operator: $ P_r \sum_{f=0}^\infty c_f / y_1^f =
c_r / y_1^r $.\\

There are only notational reasons to introduce infinite sums. For all sums in
this equation an
upper bound can be given easily. The vertex factor of the
free energy of order $ h $ can be determined with this equation
recursively. Note that only differential operators and simple projection
operators are used in this process, but no integration has
to be performed. Therefore no operator inversion in any sense is necessary to
solve the equation recursively. In that sense the equation is remarkably simple.

Usually the explicit solutions of loop equations are given up to the second
order correction ($ \sim 1 / N^4 $) (eq. (4.8) (alias eq. (54)) in
\cite{CE-06} or eq. (4-49) in \cite{EO-07}). The solution of eq. (\ref{eq:
  Lagrangean loop eq}) is written down up to the fifth order correction ($
\sim 1 / N^{10} $) in appendix A. Even higher order corrections up to the
tenth order ($ \sim 1 / N^{20} $) were calculated (table \ref{tab: term number} in appendix A) but could not be displayed due to the large number of terms.

\section{Tree Graphs}
\begin{figure}[h!] 
\begin{center}
\hspace{1.1cm}
\scalebox{0.8}{\input{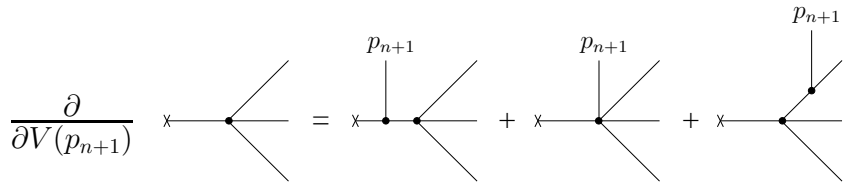} }
\caption{General step} 
\label{fig: graphical alternatives}
\end{center}
\end{figure}

The iterative application of the
loop differential operator will be seen to give rise to the combinatorial
structure  of the classical solution of the scalar theory (in a generating
external field). \\
Let a classical self-interacting scalar field theory in contact with an
external field be specified by a quadratic $ \mathcal{L}_0 (\varphi) =
\varphi D \varphi $, ($ D $ being a second order differential operator) and an interaction part $
\mathcal{L}_{int} (\varphi)  =  \mathcal{L}^{(0)} $. The action\footnote{We will finally replace the integral (of arbitrary dimension) by a discrete sum. However, this is not relevant to the arguments presented in this section.}, including the coupling to a source, is then given by 
\begin{displaymath}
S( \varphi, J ) = S ( \mathcal{L}_0 + \mathcal{L}_{int} + J \cdot \varphi ).
\end{displaymath}
After the selection of an appropriate Green's function for the differential
operator $ D $, 
\begin{displaymath}
D_x G (x,y ) = \delta^d (x-y), 
\end{displaymath}
one is lead to the zeroth order approximation of the solution of the scalar
field equation of motion
\begin{displaymath}
\varphi_{(0)} (J)(x) = \int d^d y G(x,y) J(y) \equiv  J (x) * G (x)
\end{displaymath}
and the first order approximation
\begin{equation}
\varphi_{(1)} (J)(x) =  \hat{\mathcal{L}}_{int} (\varphi_{(0)} (J)(x)) * G (x) .
\label{eq: phi_1}
\end{equation}
In $ \hat{\mathcal{L}}_{int} $, one leg of each interaction vertex of $ \mathcal{L}_{int} $ is replaced by the base point $ x $. The formal complete solution of the equation of motion $ \varphi = \varphi_{(0)} +
\varphi_{(1)} + \ldots $, to be denoted $ \varphi (J) $, is obtained through the
(infinite) iteration of eq. (\ref{eq: phi_1}). The generating functional $
G_{tree} (J) $ of all connected tree graphs of the scalar field theory is
found to be (cf. e.g. \cite{BB-68}) 
\begin{equation}
G_{tree} (J) = \int \mathcal{L}_{int} (\varphi(J)) d^d x \equiv S_{int} (\varphi (J)).
\label{eq: G_tree}
\end{equation}
To prove that the effective Lagrange density leads in fact to the same tree graph
functional as the iterated action of the loop differential operator, we first of
all note that the interaction Lagrange density is modelled to reproduce the 1-vertex
amplitudes generated via loop differentiation. The matching of the two
approaches is therefore guaranteed by construction on the level of 1-vertex
tree graphs. The completion of the argument for the tree graph approximation
is achieved according to eq. (\ref{eq: G_tree}) if it can be shown that the
formal solution of the equation of motion is generated by iterative action of
the loop operator on a Bergmann kernel (alias a Green's function of the scalar
field theory). The proof can be done by recursion in the number of external
legs. The first steps of this recursion are easily done by application of the
rules of loop differentiation. They are depicted in figure \ref{fig: loop
  differentiation}.
\begin{figure}[h!] 
\begin{center}
\hspace{1.1cm}
\epsfig{figure=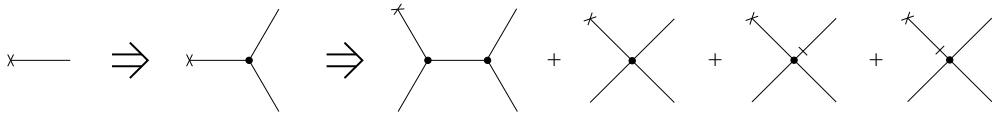, width=13cm}
\caption{First steps} 
\label{fig: loop differentiation}
\end{center}
\end{figure}
Let us now assume that to $ n $-th order (that is, for $ n $ external points
besides the base point $ x $) the assumption of equality of field theory and
matrix model amplitudes holds true. Applying $ \frac{ \partial }{ \partial V
  (p_{n+1}) } $ to that amplitude leads to the graphical alternatives
displayed in figure \ref{fig: graphical alternatives},
i.e., the newly created external line ending at $ p_{n+1} $ is inserted either
into the propagator emerging from the base point, or from the first vertex (as
seen from the base point), or inserted into one of the trees emanating from
the first vertex. The three alternatives are clearly exclusive to each other
and add up---given the induction hypothesis---to the sum of all tree graphs of
$ n+1 $ external points.

\newpage


 
\section{Induction on h}
\label{section: Induction on h}
It is assumed that the statement of theorem $ 1a $ is true for $
W_2^{(h-1)} $ and for all $ W_1^{(j)} $ with $ j=1,\ldots,h-1 $. The assertion is that the prefactors of the diagrams resulting from the loop
equation are equal to the symmetry factors of the diagrams. This assertion is assumed in the proof by induction to be true
for all $ W_1^{(j)} $, $ W_2^{(j)} $ with $ j=1,\ldots,h-1 $. The main idea of the proof is to avoid all complicated diagrams of $ W_1^{(h)}
$ generated in the loop equation (\ref{eq: loop eq. to order h}) by the
second, third or fourth term on the r.h.s. of eq. (\ref{eq: integral contour twist
  result}) and, instead, to prove theorem $ 1a $ directly for $
\mathcal{F}^{(h)} $. Since $ \mathcal{F}^{(h)} $ contains less diagrams than $ W_1^{(h)} $, only some diagrams of $ W_1^{(h)} $ generated by the first term on the r.h.s. of eq. (\ref{eq: integral contour twist
  result}) have to be inspected to obtain all diagrams of $ \mathcal{F}^{(h)} $ (with the exception of the diagram with one vertex of topological
index $ h $).\\

In the following, the field theory method for calculating symmetry factors is
introduced. Let $ m_\mu (D) $ be the multiplicity of vertices with structure $ \mu
$ in diagram $ D $. Let $ R (D) $ be the set of all vertex structures in a
diagram $ D $. $ \alpha_i (\mu) $ is the number of emanating lines with a $ i
$-th derivative in a vertex with the structure $ \mu $. The inverse symmetry factor $ \Pi_D $ can be calculated by 
\begin{displaymath}
 \Pi_D = \frac{ 1
}{ c_D  d_D } 
\ \pi_D
\end{displaymath}
\begin{displaymath}
\mbox{with\ \ \ } c_D = \prod\limits_{ \mu \in R(D) }  m_\mu (D) !,  
\mbox{\ \ \ \ \ \ \ \ \ \ \ \ \ } d_D = \prod\nolimits_{ j=1 }^n  \alpha ( \mu_j ) !
\end{displaymath}
\begin{eqnarray}
\mbox{and\ \ \ } \pi_D = \ \left< 0 \left| \p (p_1) \ldots \p (p_k) \sum_{ i_1 ,\ldots ,i_n = 1
}^{2s} \prod_{ j = 1 }^n \left( \lambda^{(h_j)}_{ \alpha( \mu_j ) , i_j }
 \p^{
  \alpha ( \mu_j ) }_{ i_j }  \right) \right| 0 \right>_D \ \ \frac{ 1 }{  S_D }.
\label{eq: pi_D}
\end{eqnarray}
$ \langle 0 | \ldots | 0 \rangle_D $ only contains the part of $ \langle 0 |
\ldots | 0 \rangle $ belonging to diagram $ D $. $ \pi_D $ as defined above denotes the number of possible pairings of the
fields in eq. (\ref{eq: pi_D}) which lead to diagram $ D $. Within the diagram $
D $, one edge could be equivalent to another edge and therefore the number of
equivalence classes of edges in $ D $ could be smaller than the number of
edges. \\

\begin{figure}[h!] 
\begin{center}
\hspace{1.1cm}
\scalebox{0.9}{\input{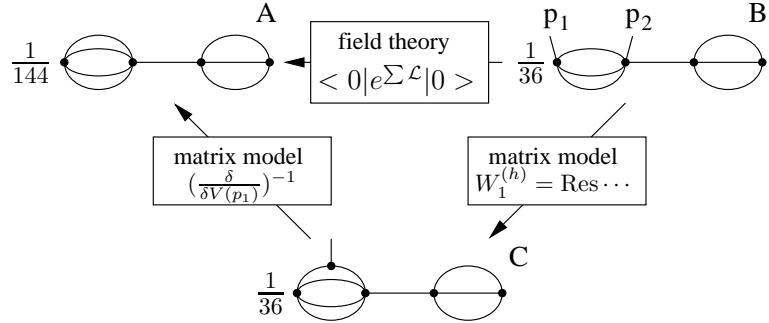} }
\caption{Example 1 for determining the prefactor of diagram $ A $. All vertices are of topological index $ 0 $.} 
\label{fig: example h-1 to h}
\end{center}
\end{figure}

An example of the method used to assure that the field theory produces the same weights of the diagrams as those produced by the matrix model is given in figure \ref{fig: example h-1 to h}.

\begin{figure}[h!] 
\begin{center}
\hspace{1.1cm}
\scalebox{0.9}{\input{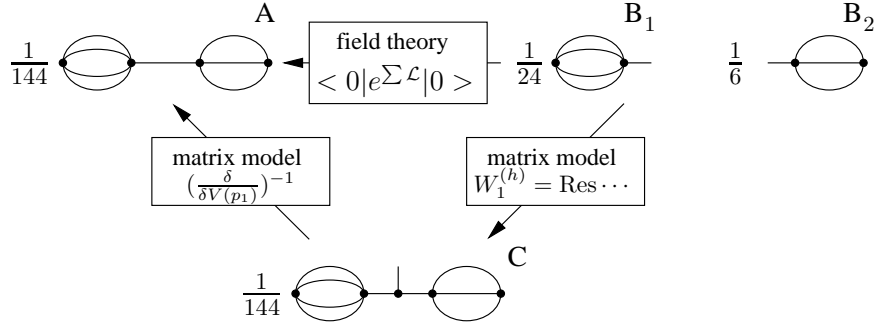} }
\caption{Example 2 for determining the prefactor of diagram $ A $ for the case where the diagram splits in two parts.} 
\label{fig: example_case2}
\end{center}
\end{figure}

Consider the diagram $ A $ from $ \mathcal{F}^{(h)} = \mathcal{F}^{(5)} $, which is depicted on the l.h.s. of figure \ref{fig: example h-1 to h}. Its prefactor, $ \frac{1}{144} $, is determined via the matrix model method and via the field theory method. One of the several diagrams or diagram pairs from $ W_1^{(4)} $ corresponding to diagram $ A $ is chosen by cutting one edge of $ A $. This diagram, where $ p_1 $ is attached to the left free end and $ p_2 $ to the right free end, is denoted $ B $. The diagram resulting from the interchange of $ p_1 $ and $ p_2 $ in $ B $ is denoted $ B^\prime $. A calculation of the symmetry factor of the diagram by the field theory method, i.e. counting the number of possible Wick contractions of the expression 
\begin{equation}
\Big< 0 \Big| \p (p_1) \p (p_2) e^{\sum_{i=1}^{2s}  \mathcal{L}(a_i)} \Big| 0 \Big>
\end{equation}
leading to diagram $ B $, results in $ \frac{1}{36} $. By the induction assumption, the matrix model correlator at order $ h-1 =  4 $ has the same prefactor. The application of the loop equation from the matrix model in form of eqs. (\ref{eq: loop eq. to order
h}) and (\ref{eq: integral contour twist result}) to the contributions of $ B $ and $ B^\prime $ leads to a prefactor of $ \frac{1}{36} $ of the diagram $ C $ from $ W_1^{(4)} $. By taking into account the fact that the loop operator can act on $ 4 $  equivalent edges in $ A $ to obtain $ C $, the prefactor of $ A $ is determined to be $ \frac{1}{144} $ by the matrix model method. For the field theory method, the counting of possible Wick contractions in the expression
\begin{equation}
\Big< 0 \Big| e^{\sum_{i=1}^{2s}  \mathcal{L}(a_i)} \Big| 0 \Big>
\end{equation}
leading to diagram $ A $ gives $ \frac{1}{144} $. Another example, where a different edge of diagram $ A $ is cut, is represented in figure \ref{fig: example_case2}. The procedure demonstrated in these examples is generalised for the proof.\\
In $ \mathcal{F}^{(h)} $ there is only one diagram with a vertex of topological index $ h $ (e.g., the first diagram in the first line of figure \ref{fig: Free energy of order two}). This diagram of the field theory gives a priori by eq. (\ref{eq: Lagrangean loop eq}) the correct matrix model contribution.\\

 Consider a diagram $ A $ from $ \mathcal{F}^{(h)} $ which does not contain a vertex of topological index $ h $. Choose one edge of $ A $ which is cut.\\

{\bf Example 1}\\
{\it The cutting procedure produces one diagram $ B $ from $ W_2^{(h-1)} $ which is
transformed into $ A $ by connecting the two external legs.}\\

\begin{figure}[h!] 
\begin{center}
\hspace{1.1cm}
\scalebox{0.4}{\input{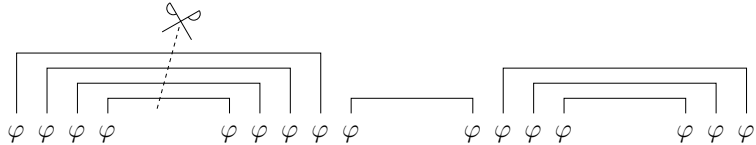} }
\caption{One contraction of $ A $ corresponds to four contractions of $ B $, i. e. $ \pi_B = 4 \pi_A $. Each of the four equivalent edges can be cut and replaced by contracting the left open end to $ \p (p_1) $ and the right open end to $ \p (p_2) $.} 
\label{fig: contra1}
\end{center}
\end{figure}

The prefactor $
\Pi_B $ of diagram $
B $ is known according to the induction assumption. It is possible that the exchange of external legs creates a second diagram $ B^\prime $ which is different to $ B $. In this case, $ u = 2 $. However, if $ B^\prime $ is identical to $ B $, then $ u = 1 $. The loop equation (eq. (\ref{eq: loop eq. to order
h})) is applied to diagram $ B $ (and diagram $ B^\prime $). Then the first term on the r.h.s. of eq. (\ref{eq: integral contour twist result}), $
1/2 \cdot B_{j,i}^{f,0} \cdot ( B_i^0 (p) / y_{1,i} ) \cdot B_{i,k}^{0,g} $, gives rise to a diagram $ C $ from $ W_1^{(h)} $. Since $ C $ cannot be created from other diagrams and, in the recursion
formula, a multiplication with $ 1/2 $ occurs, the prefactor of $ C $ is given
by $ \Pi_C = \frac{ 1 }{ 2 } u \Pi_B $.\\ 
The external leg of $ C $ is rooted in a vertex of index zero with 3 emanating
lines. $ C $ can only have been created by $ \Delta_7 $ from a diagram of $
\mathcal{F}^{(h)} $. This diagram must have been $ A $. Let $ v $ be the number of edges in the equivalence class of the edge which is cut. Then $ C $ can be created by $ \Delta_7 $ from $ A $ in $ v $
different ways. The matrix model calculation yields $ \Pi_C = v \Pi_A $ and hence $ \Pi_A = \frac{
  u }{ 2 v } \Pi_B $.\\
In the following, the quotient $ \Pi_A / \Pi_B $ is investigated in the
field theory language. The specific example of figure \ref{fig: example h-1 to h} is generalised for the proof. In figure \ref{fig: contra1} one special contraction of diagram $ A $ is given and compared to four corresponding special contractions from diagram $ B $.

In this way, each of the $ \pi_A $ contractions of $ A $ corresponds to $ 4 $ contractions of $ B $: $ \pi_B = 4 \pi_A $. The generalisation of this formula for $ v $ equivalent edges in $ A $ is $ \pi_B = v \pi_A $.\\
If the interchange $ \p (p_1) \leftrightarrow \p (p_2) $ in $ B $ does not give a new diagram $ B^\prime $ different from  $ B $, then for every contraction in $ A $ there are two in $ B $: $ \pi_B = 2 \pi_A $. Combining both equations leads to $ \pi_B = \frac{2}{u} v \pi_A $. $ c_B  = c_A $ and $ d_B = d_A $ imply that $ \Pi_B = \frac{1}{c_B d_B} \pi_B =  \frac{1}{c_A d_A} \frac{2}{u} v \pi_A = \frac{2}{u} v \Pi_A $.\\ 

{\bf Example 2}  \\ 
{\it The cutting procedure produces two diagrams $ B_1 $ and $ B_2 $.} \\

There is a $ j \in \{1,\ldots,h-1\} $, such that $ B_1 $ is a diagram
from $ W_1^{(j)} $ and $ B_2 $ is a diagram from $ W_1^{(h-j)} $. The loop equation (eq. (\ref{eq: loop eq. to order
h})) is applied to the diagrams $ B_1 $ and $ B_2 $. Then the first term on the r.h.s. of eq. (\ref{eq: integral contour twist result}), $
1/2 \cdot B_{j,i}^{f,0} \cdot ( B_i^0 (p) / y_{1,i} ) \cdot B_{i,k}^{0,g} $, gives rise to a diagram $ C $ from $ W_1^{(h)} $. The diagram $ C $ does not receive additional contributions from $ W_2^{(h-1)} $ in eq. (\ref{eq: loop eq. to order
h}) for the following reasons. The diagrams generated by the first term on the r.h.s. of eq. (\ref{eq: integral contour twist result}) do not break in two parts in the cutting procedure, whereas the other terms on the r.h.s. of eq. (\ref{eq: integral contour twist result}) produce a non-zero topological index at the external leg.\\
For $ j \neq h/2 $, the diagram $ C $ appears twice in the loop equation
summation $
\sum_{m=1}^{h-1} W_1^{(m)} W_1^{(h-m)} $. For $ j=h/2 $ and $ B_1 \neq B_2 $,
the situation does not change. However, for $ B_1 = B_2 $, the diagram $ C $ only
appears once. Let $ u = 2 $ for $ B_1 \neq B_2 $ and $ u =1 $ for $ B_1 =
B_2 $. Since in the first term of the recursion formula (eq. (\ref{eq:
  integral contour twist result})) one multiplies with
$ 1/2 $, this gives $ \Pi_C = \frac{ u }{ 2 } \Pi_{B_1} \Pi_{B_2} $. The
external leg of $ C $ is connected to a vertex of topological index zero with
3 emanating lines. $ C $ can only have been created by the action of $
\Delta_7 $ on a diagram of $ \mathcal{F}^{(h)} $. This diagram must be diagram
$ A $. Let $ v $ be the number of elements in the equivalence class of the edge which is cut. Then $ C $ can be deduced from $ A $ in $ v $ different ways. This matrix model calculation yields
$ \Pi_C = v \Pi_A $ and hence $ \Pi_A = \frac{ u }{ 2 v } \Pi_{B_1}
\Pi_{B_2} $.\\
In the following, the quotient between $ \Pi_A $ and $ \Pi_{B_1} \Pi_{B_2} $ is
investigated in the field theory language.

The specific example of figure \ref{fig: example_case2} is generalised for the proof. 

 Since in $ A $ two vertices of the same vertex structure (topological index zero and four emanating edges without derivatives) appear and only one is part of $ B_1 $, there are $ \left( { 2 \atop 1 } \right) $ more contraction possibilities in $ B $ than in $ B_1 $ and $ B_2 $. In general, this factor is $ \Pi_{ \mu \in R(A) } \left( { m_\mu (A) \atop m_\mu (B_1) } \right) $. If $ B_1 = B_2 $ then this number is reduced by $ \frac{1}{2} $ because the assignment of vertices to $ B_1 $ gives the same contractions in $ A $ as the assignment of the complementary vertices to $ B_1 $. If the edge which is cut in $ A $ is one of $ v $ equivalent edges in $ A $ then the number of contractions $ \pi_{B_1} \cdot \pi_{B_2} $ is enhanced by a factor $ v $. This gives in total $ v \pi_A = \frac{u}{2}  \Pi_{ \mu \in R(A) } \left( { m_\mu (A) \atop m_\mu (B_1) } \right)  \pi_{B_1} \pi_{B_2} $. The relation of the Gell-Mann-Low factorials is $ \frac{ c_A }{ c_{B_1} c_{B_2} }= \Pi_{ \mu \in R(A) } \left( { m_\mu (A) \atop m_\mu (B_1) } \right) $ and hence $ 
\Pi_A = \frac{u}{2 v}  \Pi_{B_1} \Pi_{B_2} $.\\

This closes the proof by induction on $ h $.\\

\section{Induction on k}
\label{section: Induction on k}
It is assumed that the statement of theorem $ 1a $ is correct for a certain $ h
$ and $ k $. In this section, we show that theorem $ 1a $ is also true for $ h $ and $ k+1 $.\\

To prove the induction step from $ k $ to $ k+1 $, two assertions have to hold:
the prefactor of any diagram $ A $ from $ W_{k+1}^{(h)} $ arising from the action
of the loop operator must be equal to the inverse symmetry factor of diagram $
A $ which is determined by counting Wick contractions. Secondly, the vertex
factor  emerging at the new vertex from the action of the loop operator must
be equal to the coupling constant of the field theory defined in
eq. (\ref{eq: lambda}). Provided the induction assumption holds for any
diagram $ B $ from $ W_k^{(h)} $, the former assertion is true if the ratio $
\Pi_A / \Pi_B $ calculated with the field theory method is equal to the ratio
$ \Pi_A / \Pi_B $ calculated with the loop operator method.\\ 
 Let $ A $ be a diagram
of $ W_{k+1}^{(h)} $.\\ 

{ \bf Case 1}\\
{\it The leg $ p_{k+1} $ emanates in diagram $ A $ from a vertex with a vertex structure $ \mu = ( 0, (3,0)) $.}
\\
\\
Only the action of $ \Delta_7 $ or $ \Delta_4 $ on diagrams of $ W_{k}^{(h)} $ with
the same number of loops as in $ A $ and one vertex less than $ A $ can have generated this diagram. The
multiplicity of the edge in the preimage diagram $ B $, which is cut, is denoted $
v $, i.e. the equivalence class of such an edge in $ B $ contains $ v $
elements. Let $ m_A $ be the number of vertices with $ \alpha = (3,0) $ and
topological index $ 0 $ in\nolinebreak$\ $\nolinebreak$ A $.

The number of contractions $ \pi_A $ in diagram $ A $ is larger than the number of contractions $ \pi_B $ in diagram $ B $. One of the $ m_A $ vertices can be chosen for the new vertex, then $ 3 $ possibilities exist to choose one leg of that vertex for the connection to $ \p (p_{k+1}) $ and $ 2 $ more possibilities to assign the remaining two free legs of the new vertex to the contraction which is cut in diagram $ B $. Eventually, since the new vertex can be inserted in each of the $ v $ equivalent individual contractions of the considered complete contraction of $ B $ we have
\begin{equation}
\frac{ \pi_A }{ \pi_B } = m_A \cdot 3 \cdot 2 \cdot v   .
\end{equation}
The quotient of the Gell-Mann-Low factorials $ c_B / c_A = 1 / m_A $ and the quotient of the factorials from the definition of the Lagrange density $ d_B / d_A = 1 / 3! $ result in 
\begin{equation}
 \Pi_A = v \Pi_B, 
\end{equation}
which is also predicted by the action of $ \Delta_7 $ or $ \Delta_4 $ in the matrix model calculation.\\ 
The emerging vertex factor at the new vertex is $ 1 / y_1 $, which is exactly
the definition of $ \lambda_{(3,0)}^{(0)} $ in eq. (\ref{eq: lambda}).\\

{ \bf Case 2}\\
{\it The leg $ p_{k+1} $ has at least one derivative.}\\ 

Only the action of $ \Delta_2 $ on diagrams of $ W_{k}^{(h)} $ with
the same number of loops as in $ A $ and the same number of vertices can have
generated the diagram $ A $. The structure of the vertex connected to $
p_{k+1} $ is denoted $ \mu_A = ( h_0, \alpha ) $ and the number of derivatives
at the leg $ p_{k+1} $ is $ r > 0 $. Let $ m_A 
$ be the
number of vertices with structure $ \mu_A $ in $ A $. Let $ m_B $ be the number of preimage vertices in $
B $ with the same structure, of which $ v $ are equivalent.\\

The number of contractions $ \pi_A $ in diagram $ A $ is compared to the number of contractions $ \pi_B $ in diagram $ B $. The interchange of one of the $ m_B $ vertices in $ B $ with the structure of the preimage vertex leads to $ m_B $ different complete contractions which give rise (for $ m_A = \alpha_r = v = 1 $) to only one contraction for diagram $ A $. Similarly, if $ m_A > 1 $ then the number of contractions in $ A $ is reduced by the factor $ m_A $ in comparison to the contractions of $ B $. Additionally, the number of contractions in $ A $ is enhanced by choosing one of $ v $ equivalent vertices and one of $ \alpha_r $ legs for $ p_{k+1} $:
\begin{equation}
\frac{ \pi_A }{ \pi_B } = \frac{ m_A  }{ m_B } \cdot \alpha_r \cdot v  .
\end{equation}
The quotient of the Gell-Mann-Low factorials $ c_B / c_A = m_B / m_A $ and the quotient of the factorials from the definition of the Lagrange density $ d_B / d_A = 1 / \alpha_r $ result in 
\begin{equation}
 \Pi_A = v \Pi_B, 
\end{equation}
which is also predicted by the action of $ \Delta_2 $ in the matrix model calculation.\\ \\
The vertex factor of the new vertex emerges from the action of $ \Delta_2 $ on
the diagram with the old vertex. Since in the definition of $ \lambda_\alpha^{(h_0)} $ the remnants
of $ \Delta_1 $ and $ \Delta_2 $ are used, this gives the correct vertex
factor of eq. (\ref{eq: lambda}).\\

{\bf Case 3}\\  
{\it The leg $ p_{k+1} $ in diagram $ A $ has no derivative and is rooted at a vertex with vertex structure $ \mu \neq ( 0 , ( 3, 0 ) ) $.}\\


In contrast to the other cases, in this case, the new diagram emerges from
different parts ($ \Delta_1 $, $ \Delta_3 $ and $ \Delta_{5/6} = \Delta_5 + \Delta_6 $) of
the loop operator. Therefore the assertion that the new vertex factor is
described by eq. (\ref{eq: lambda}) is closely related to the different symmetry
factors of the preimage diagrams.\\
There is at most one preimage diagram which becomes diagram $ A  $ after action of $
\Delta_1 $. This is denoted as $ B_0 $. Each external or internal line (apart from the leg $ p_{k+1} $ itself) emanating from the
vertex connected with $ p_{k+1} $ gives possibly one further preimage diagram,
which transforms under the action of $ \Delta_3 $ or $ \Delta_{5/6} $ to diagram $ A $. In $ A $, the outgoing external and internal legs emanating from the vertex connected to $
p_{k+1} $ are named arbitrarily, with $ f = 1, \ldots, n $, and their
derivatives are denoted as $ r_1, \ldots, r_n $ respectively. \\
If $ r_f = 0  $ then no preimage diagram $ B_f $ associated to leg $ f $
exists. This kind of `missing diagram' does not present large difficulties in
the computation of the new vertex factor of $ A $, because in the
consideration of a vertex consisting exclusively of external legs, $ \Delta_3 $
can not act on such a leg without derivative as $ \Delta_{5/6} $ can not act
on $ f $. To describe the second kind of `misssing diagrams' we define $ \delta $ to be the number of diagrams from $ 
 B_1 , \ldots, B_n
 $ which are equal to $ B_f $. If $ \delta > 1 $, there is only one preimage diagram for $ \delta $ legs, rather than one for each leg. \\

The following notation will be used.
The number of vertices in $ A $ with the same structure as the vertex
connected to $ p_{k+1} $ is denoted $ m_A $. The structure of the vertex
connected to $ p_{k+1} $ is defined as $ \mu_A = ( h_0, \alpha ) $ and $ \gamma $ is defined as the number of possible ways to apply $ \Delta_5 $ or $ \Delta_3 $ to $ B_f $ to
obtain $ A $.\\ 
 
{\bf Symmetry factors of the diagrams $ B_0 $ and $ A $:}\\
Let the number of possible ways to apply $ \Delta_1 (p_{k+1}) $ to $ B_0 $
to obtain $ A $ be $ v $. This is the number of equivalent preimage vertices. Let $ m_B $ be the number of vertices with the structure
$ ( h_0, ( \alpha_0 - 1 , \alpha_1, \alpha_2, \ldots ) ) $ in $ B_0 $.

The number of contractions $ \pi_A $ in diagram $ A $ is compared to the number of contractions $ \pi_{B_0} $ in diagram $ B_0 $. The interchange of one of the $ m_B $ vertices in $ B $ with the structure of the preimage vertex leads to $ m_B $ different complete contractions which give rise (for $ m_A = \alpha_0 = v = 1 $) to only one contraction for diagram $ A $. Similarly, if $ m_A > 1 $ then the number of contractions in $ A $ is reduced by the factor $ m_A $ relative to the number of contractions of $ B_0 $. Additionally, the number of contractions in $ A $ in enhanced by choosing one of $ v $ equivalent vertices and one of $ \alpha_0 $ legs for $ p_{k+1} $:
\begin{equation}
\frac{ \pi_A }{ \pi_{B_0} } = \frac{ m_A }{ m_B } \cdot \alpha_0  \cdot v  .
\end{equation}
The quotient of the Gell-Mann-Low factorials $ c_B / c_A = m_B  / m_A $ and the quotient of the factorials from the definition of the Lagrange density $ d_B / d_A = 1 / \alpha_0 $ result in 
\begin{equation}
\label{eq: field theory quotient B_0}
 \Pi_A = v \Pi_{B_0} 
\end{equation}
which is also predicted by the action of $ \Delta_1 $ in the matrix model calculation.\\

{\bf Symmetry factors of the diagrams $ B_f $ with $ f > 0 $ and $ A $:}\\
Let $ v_A $ be the number of elements in the equivalence class of edge $ f $ in
$ A $. The number of derivatives at edge $ f $ in diagram $ A $ is $
r $ (also denoted $ r_f $) at the
end of edge oriented towards $ p_{k+1} $. The vertex structure of the vertex connected to $ p_{k+1} $ is denoted $
\mu_A $ and the structure of the preimage vertex in $ B_f $ is $ \mu_B = ( h_0 , \beta ) $ with $ \beta_i = \alpha_i - \delta_{i,0} + \delta_{ i, r-1 } - \delta_{ i,r } $. Then we define $ m_A = m_{\mu_A} (A) $ and $ m_B = m_{\mu_B} (B_f) $. Let $ \hat{f} $ be one edge in $ B_f $ with $ r_f - 1 $ derivatives which
transforms under the action of $ \Delta_3 $ or $ \Delta_{5/6} $ to the edge $ f $ in diagram $
A $. The number of elements in the equivalence class of edge $ \hat{f} $ in $
B_f $ is denoted $ v_B $.\\

The number of contractions $ \pi_A $ in diagram $ A $ is compared to the number of contractions $ \pi_B $ in diagram $ B_f $. The vertex connected to $ p_{k+1} $ in diagram $ A $ is one out of a total of $ m_A $ vertices. In diagram $ B_f $, this vertex is replaced by a vertex with the structure $ \mu_B $, which is one out of a total of $ m_B $ vertices. From the choice of the vertex in $ A $ or $ B_f $ the factor $ m_A / m_B  $ appears in the relation of $ \pi_A $ to $ \pi_B $. At this vertex, the leg $ p_{k+1} $ can be connected to one of $ \alpha_0 $ legs without derivatives, which gives rise to the factor $ \alpha_0 $ in the comparison of the contractions. For the same reason, the connection of the edge $ f $ to the vertex leads to a factor $ \alpha_r $. On the other hand, the connection of the edge $ \hat{f} $ to the preimage vertex with the structure $ \mu_B $ leads to a factor $ 1 / \beta_{r-1} $.\\

If the edge $ f $ in $ A $ is equivalent to another edge in $ A $ or the edge $ \hat{f} $ is equivalent to another edge in $ B_f $ or the edge $ f $ connects a vertex with itself then the relation $ \gamma / \delta $ is also part of the comparison of contractions. $ \gamma $, as can be read from the definition, is equal to $ v_B $ or $ 2 v_B $, where the latter occurs if and only if the edge $ f $ connects a vertex with itself and the number of derivatives at the other end of this edge is equal to $ r-1 $. $ \delta $, as can be read from the definition, is equal to $ v_A $ or $ 2 v_A $, where the latter occurs if and only if edge $ f $ connects a vertex with itself and the number of derivatives at the other end of this edge is equal to $ r $. The enlargement of the number of equivalent edges by a factor $ 2 $ in these special examples reflects the fact that, in the contractions for each equivalent edge, $ 2 $ legs instead of $ 1 $ leg of that vertex are affected. This gives, in total,
\begin{equation}
\frac{ \pi_A }{ \pi_{B_f} } =  \frac{ m_A }{ m_B } \frac{ \alpha_0 \alpha_r }{  \beta_{r-1}  }  \frac{ \gamma }{ \delta }  
.
\end{equation}
The quotient of the Gell-Mann-Low factorials $ c_B / c_A = m_B / m_A $ and the quotient of the factorials from the definition of the Lagrange density $ d_B / d_A = \beta_{r-1} / (\alpha_0 \alpha_r) $ result in 
\begin{equation}
\label{eq: field theory quotient B_f}
 \Pi_A = \frac{ \gamma }{ \delta } \Pi_{B_f} .
\end{equation}
In the loop operator determination of the relation of $ \frac{ \Pi_{ B_f
  } }{ \Pi_A } $, one has to consider, on the one hand, that there
are $ \gamma $ possible ways to apply $ \Delta_3 $ or $ \Delta_{5} $ to the
diagram $ B_f $ and on the other hand that the number of diagrams in comparison to a one-vertex
calculation with exclusively external legs is lowered by a factor $ 1 / \delta $. The matrix model calculation therefore yields 
\begin{equation}
\label{eq: mm quotient B_f}
 \Pi_A = \frac{ \gamma }{ \delta } \Pi_{B_f}, 
\end{equation}
which coincides with the field theory calculation from eq. (\ref{eq: field theory quotient B_f}). \\ 

{\bf New vertex factor}\\
The definition of the vertex factor is based on $ \Delta_1 $ and $ \Delta_2 $,
but it was shown in (\cite{FGK-08}, page 12) that the vertex factors of the
diagrams with only 1 vertex without internal lines are correctly described by
eq. (\ref{eq: lambda})---this includes the action of $ \Delta_3 $. We compare the
action of the loop operator on the preimage diagrams leading to the vertex
connected to $ p_{k+1} $ in $ A $ to the case in which the loop operator acts on a vertex of the same structure with solely external legs. No difference occurs for the action of $ \Delta_3 $ to the external lines. The action of $
\Delta_{5/6} $ on the internal lines can be regarded as an action of $ \Delta_3
$ on additional external lines. The contribution of $ B_0 $ to the new
vertex factor is---relative to the corresponding 1-vertex diagram---enhanced
by a factor $ v \Pi_{ B_0 } $. The contribution of $ B_f $ to the new vertex
factor is---relative to the corresponding contribution to the 1-vertex
diagram---enhanced by a factor $ \gamma \Pi_{ B_f } $, where $ \gamma $ is
equal to the number of possible ways to apply $ \Delta_3 $ or $ \Delta_5 $ to $ B_f $ to
obtain $ A $. Instead of $ \delta $ preimage diagrams for the 1-vertex diagram
there is only one. Consequently, the new vertex factor arises in this complicated vertex in exactly the same way as in 1-vertex
diagrams if the equation $ v \Pi_{B_0}  =  \gamma \Pi_{B_f} / \delta $ is
valid. The combination of eqs. (\ref{eq: field theory quotient B_0}) and (\ref{eq: mm quotient B_f}) from the previous paragraph establishes this condition and hence the new vertex is described by eq. (\ref{eq: lambda}).\\

This closes the proof by induction on $ k $.

Starting from $ W_3^{(0)} $ and $ W_1^{(1)} $ (shown in \cite{FGK-08} to be
described by the Lagrangean formalism) one
applies the induction on $ k $ from section four to prove that theorem $ 1a $ holds for $ W_k^{(0)} $ with $ k \ge 3 $ and $ W_k^{(1)} $ with $ k \in
\mathbb{N} $. From the
application of the induction on $ h $, it follows that theorem $ 1a $ also
holds for $ W_0^{(2)} = \mathcal{F}^{(2)} $. The application of the induction on $
k $ reveals that theorem $ 1a $ holds for all $ W_k^{(2)} $ with $ k \in \mathbb{N}_0 $. The
$ (h-2) $-fold alternating use of the inductions on $ h $ and $ k $  proofs
that $ W_k^{(h)} $ with $ h \ge 3 $ and $ k \in \mathbb{N}_0 $ is described by
a field theory. This concludes the proof of theorem $ 1a $, and, thereby, the proof of theorem $ 1 $.

\section{Summary}
\label{section: Summary}
In this article we have proved that the Lagrangean formulation of the
Hermitean 1-matrix models, introduced in \cite{FGK-08}, is valid to all orders
in the genus expansion. This was achieved by comparing the prefactors of all
terms arising from the loop operator with the symmetry factors of the
corresponding diagrams and finding full agreement. The free energy was directly incorporated into the statement of the theorem. Furthermore, the procedure to obtain coupling constants of a given topological
index from the coupling constants of lower indices, which was solved up to third
order in \cite{FGK-08}, was condensed into one
equation. This is the loop equation of the Lagrangean formalism.\\ It was solved
to the tenth order and explicit expressions were given to the fifth
order (See appendix A). A new method of determining the free energy from the one-point function was
found by inverting a part of the loop operator giving rise to a simple
integration. This new method does not, contrary to \cite{FGK-08}, rely on
results for the properties of the operator $ H $ obtained in \cite{CE-06}.\\

{\large \bf Acknowledgments} \\ 
The author would like to thank R. Flume and J.B. Zuber for discussions, and K.E. Williams for a careful reading of the manuscript.

\newpage
{\LARGE \bf Appendix A}\\

This appendix contains the explicit solution of the loop equation for the
coupling constant to the orders three to five. The lower orders were
already given in section \ref{section: Lagrangean formalism}. The orders six to ten were
also calculated, but are not displayed due to the large number of terms (see
table \ref{tab: term number}).

\begin{table}[h]
\begin{tabular}{|r||c|c|c|c|c|c|c|c|c|c|c|}  \hline
h & 0 & 1 & 2 & 3 & 4 & 5 & 6 & 7 & 8 & 9 & 10 \\ \hline
\# of terms in $ \lambda^{(h)} $ & 1 & - & 3 & 11 & 30 & 77 & 176 & 385 & 792 &
1575 & 3010 \\ \hline
\end{tabular}
\caption{number of terms in $ \lambda^{(h)} $}
\label{tab: term number}
\end{table}

\begin{eqnarray*}
\lambda^{(3)}=  
 \frac{2205}{256} \frac{y_2^6}{y_1^{10}}
-\frac{8685}{256} \frac{y_2^4 y_3}{y_1^9}
+\frac{15375}{512} \frac{y_2^2 y_3^2}{y_1^8}
+\frac{5565}{256} \frac{y_2^3 y_4}{y_1^8}
-\frac{72875}{21504} \frac{y_3^3}{y_1^7} 
 \\ 
-\frac{5605}{256} \frac{y_2 y_3 y_4 }{y_1^7}
-\frac{3213}{256} \frac{y_2^2 y_5}{y_1^7}
+\frac{21245}{9216} \frac{y_4^2}{y_1^6}
+\frac{2515}{512} \frac{y_3 y_5}{y_1^6}
+\frac{5929}{1024} \frac{y_2 y_6}{y_1^6}
-\frac{5005}{3072} \frac{y_7}{y_1^5}
\end{eqnarray*}

\begin{eqnarray*}
\lambda^{(4)} =
-{\frac {8437275}{32768}}    \frac{ y_5y_6 }{ y_1^8 }       
+{\frac {1511055}{2048}}      \frac{ y_2y_5^2 }{ y_1^9 } 
-{\frac {12677665}{32768}}   \frac{ y_2y_9 }{ y_1^8 }  
-{\frac {32418925}{24576}}   \frac{ y_3^3y_4 }{ y_1^{10} } \\ 
+{\frac {11532675}{16384}}    \frac{ y_3^2y_6 }{ y_1^9 } 
-{\frac {10156575}{32768}}   \frac{ y_3y_8 }{ y_1^8 } 
-{\frac {8913905}{32768}}    \frac{ y_4y_7 }{ y_1^8 } 
+{\frac {4456305}{512}}       \frac{ y_2^6y_4 }{ y_1^{13} } \\ 
-{\frac {12829887}{1024}}    \frac{ y_2^7y_3 }{ y_1^{14} } 
+{\frac {98342775}{4096}}     \frac{ y_2^5y_3^2 }{ y_1^{13} } 
-{\frac {12093543}{2048}}    \frac{ y_2^5y_5 }{ y_1^{12} } 
+{\frac {15411627}{4096}}     \frac{ y_2^4y_6 }{ y_1^{11} } \\ 
-{\frac {16200375}{1024}}    \frac{ y_2^3y_3^3 }{ y_1^{12} } 
-{\frac {44207163}{20480}}   \frac{ y_2^3y_7 }{ y_1^{10} } 
+{\frac {83895625}{32768}}    \frac{ y_2y_3^4 }{ y_1^{11} } 
-{\frac {578655}{128}}       \frac{ y_2y_3^2y_5 }{ y_1^{10} } \\ 
+{\frac {13024935}{1024}}     \frac{ y_2^3y_3y_5 }{ y_1^{11} } 
+{\frac {12367845}{2048}}     \frac{ y_2^3y_4^2 }{ y_1^{11} }  
+{\frac {21023793}{10240}}    \frac{ y_2^9 }{ y_1^{15} } 
-{\frac {26413065}{1024}}    \frac{ y_2^4y_3y_4 }{ y_1^{12} } \\ 
+{\frac {7503125}{36864}}     \frac{ y_4^3 }{ y_1^9 } 
+{\frac {4297293}{4096}}      \frac{ y_2^2y_8 }{ y_1^9 } 
+{\frac {2642325}{2048}}      \frac{ y_3y_4y_5 }{ y_1^9 } 
-{\frac {5472621}{1024}}     \frac{ y_2^2y_3y_6 }{ y_1^{10} } \\ 
-{\frac {10050831}{2048}}    \frac{ y_2^2y_4y_5 }{ y_1^{10} } 
+{\frac {8083075}{98304}}     \frac{ y_{10} }{ y_1^7 } 
-{\frac {17562825}{4096}}    \frac{ y_2y_3y_4^2 }{ y_1^{10} } 
+{\frac {6968247}{4096}}      \frac{ y_2y_3y_7 }{ y_1^9 } \\ 
+{\frac {68294625}{4096}}     \frac{ y_2^2y_3^2y_4 }{ y_1^{11} } 
+{\frac {6242775}{4096}}      \frac{ y_2y_4y_6 }{ y_1^9 }
\end{eqnarray*}

\begin{eqnarray*}
\T
\lambda^{(5)} = 
-{\frac {30087162105}{32768}}                       \,\frac{ y_2^2y_3y_9   }{ y_1^{12} }      
+{\frac {20308307985}{8192}}                        \,\frac{ y_2^2y_4^2y_5 }{ y_1^{13} }     
-{\frac {26135514405}{32768}}                       \,\frac{ y_2^2y_4y_8 }{  y_1^{12}}       
+{\frac {1336297095}{512}}                          \,\frac{ y_2^2y_3y_5^2 }{ y_1^{13}  } \\ \T    
+{\frac {739690835625}{32768}}                      \,\frac{ y_2^4y_3^2y_5 }{ y_1^{15}}     
+{\frac {10981859175}{512}}                         \,\frac{ y_2^4y_3y_4^2 }{  y_1^{15}  }      
-{\frac {179984279475}{32768}}                      \,\frac{ y_2^4y_3y_7 }{ y_1^{14}}     
-{\frac {357630901425}{32768}}                      \,\frac{ y_2^2y_3^2y_4^2 }{ y_1^{14} }  \\ \T   
+{\frac {98590152375}{32768}}                       \,\frac{ y_2^2y_3^2y_7 }{ y_1^{13}}     
-{\frac {24241354137}{32768}}                       \,\frac{ y_2^2y_5y_7 }{ y_1^{12}  }     
+{\frac {40226430195}{4096}}                        \,\frac{ y_2^5y_4y_5 }{ y_1^{15}}     
-{\frac {1632433564125}{32768}}                     \,\frac{ y_2^5y_3^2y_4 }{ y_1^{16}  }     \\ \T
+{\frac {350207643555}{32768}}                      \,\frac{ y_2^5y_3y_6 }{ y_1^{15}}     
+{\frac {261787425795}{8192}}                       \,\frac{ y_2^7y_3y_4 }{ y_1^{17} }     
-{\frac {9732778965}{512}}                          \,\frac{ y_2^6y_3y_5 }{ y_1^{16}}     
-{\frac {37121844375}{4096}}                        \,\frac{ y_2^3y_3^2y_6 }{ y_1^{14}}   \\ \T  
+{\frac {11830867475}{8192}}                        \,\frac{ y_2y_3y_4^3 }{ y_1^{13} }    
-{\frac {8775042255}{8192}}                         \,\frac{ y_2^{12} }{ y_1^{20}  }     
-{\frac {877489930625}{262144}}                     \,\frac{ y_2y_3^4y_4 }{ y_1^{14}}     
+{\frac {12359275929}{65536}}                       \,\frac{ y_2y_6y_7 }{ y_1^{11}  }   \\ \T  
+{\frac {3222490635}{16384}}                        \,\frac{ y_2y_5y_8 }{ y_1^{11}}     
+{\frac {33732646101}{16384}}                       \,\frac{ y_2^3y_5y_6 }{ y_1^{13}  }     
-{\frac {26915796875}{262144}}                      \,\frac{ y_3^6 }{ y_1^{14}}     
+{\frac {217019504625}{131072}}                     \,\frac{ y_2y_3^3y_6 }{ y_1^{13}  }   \\ \T  
+{\frac {12983424025}{131072}}                      \,\frac{ y_3^2y_9 }{ y_1^{11}}     
-{\frac {5553643095}{4096}}                         \,\frac{ y_2^5y_8 }{ y_1^{14}  }     
-{\frac {2109682575}{65536}}                        \,\frac{ y_4y_{10} }{ y_1^{10}}     
-{\frac {136426972815}{8192}}                       \,\frac{ y_2^3y_3y_4y_5 }{ y_1^{14}  } \\ \T    
-{\frac {12753475735}{262144}}                      \,\frac{ y_2y_{12} }{ y_1^{10}}     
+{\frac {35383072725}{16384}}                       \,\frac{ y_2^6y_7 }{ y_1^{15}  }     
-{\frac {52671571029}{16384}}                       \,\frac{ y_2^7y_6 }{ y_1^{16}}     
+{\frac {1122904568025}{32768}}                     \,\frac{ y_2^6y_3^3 }{ y_1^{17}  }  \\ \T   
+{\frac {74926833597}{16384}}                       \,\frac{ y_2^8y_5 }{ y_1^{17}}     
+{\frac {20420786925}{262144}}                      \,\frac{ y_3y_6^2 }{ y_1^{11}  }     
+{\frac {8203397345}{32768}}                        \,\frac{ y_2y_3y_{10} }{ y_1^{11}}     
-{\frac {37992975405}{65536}}                       \,\frac{ y_2y_4^2y_6 }{ y_1^{12}  }  \\ \T   
-{\frac {250975486125}{32768}}                      \,\frac{ y_2^2y_3^3y_5 }{ y_1^{14}}     
+{\frac {144886833945}{16384}}                      \,\frac{ y_2^{10}y_3 }{ y_1^{19}  }     
-{\frac {147985547535}{16384}}                      \,\frac{ y_2^6y_4^2 }{ y_1^{16}}     
+{\frac {10137152025}{4096}}                        \,\frac{ y_2^3y_3y_8 }{ y_1^{13}  }  \\ \T   
+{\frac {21786582125}{32768}}                       \,\frac{ y_3^3y_4^2 }{ y_1^{13}}     
-{\frac {3611933325}{131072}}                       \,\frac{ y_6y_8 }{ y_1^{10}  }     
-{\frac {161462697165}{32768}}                      \,\frac{ y_2^4y_4y_6 }{ y_1^{14}}     
-{\frac {9188135775}{16384}}                        \,\frac{ y_2y_4y_5^2 }{ y_1^{12}  }  \\ \T   
+{\frac {5226845715}{32768}}                        \,\frac{ y_3y_5y_7 }{ y_1^{11}}     
+{\frac {91746650625}{262144}}                      \,\frac{ y_3^4y_5 }{ y_1^{13}  }     
-{\frac {21601719825}{8192}}                        \,\frac{ y_2^3y_4^3 }{ y_1^{14}}     
+{\frac {176853471795}{32768}}                      \,\frac{ y_2^2y_3y_4y_6 }{ y_1^{13} }     \\ \T
-{\frac {2481504025}{65536}}                        \,\frac{ y_3y_{11} }{ y_1^{10}}     
+{\frac {951059690625}{262144}}                     \,\frac{ y_2^2y_3^5 }{ y_1^{15} }     
-{\frac {12688164489}{32768}}                       \,\frac{ y_2^3y_{10} }{ y_1^{12}}     
-{\frac {26018510375}{131072}}                      \,\frac{ y_3^3y_7 }{ y_1^{12}  }     \\ \T
-{\frac {19518799797}{8192}}                        \,\frac{ y_2^4y_5^2 }{ y_1^{14}}     
+{\frac {4685481675}{32768}}                        \,\frac{ y_4y_5y_6 }{ y_1^{11}  }     
+{\frac {17861648805}{8192}}                        \,\frac{ y_2^3y_4y_7 }{ y_1^{13}}     
-{\frac {16063796175}{32768}}                       \,\frac{ y_3y_4^2y_5 }{ y_1^{12} }   \\ \T  
-{\frac {47335422519}{131072}}                      \,\frac{ y_2^2y_6^2 }{ y_1^{12}}     
-{\frac {625336273125}{32768}}                      \,\frac{ y_2^4y_3^4 }{ y_1^{16}}     
-{\frac {430379379675}{16384}}                      \,\frac{ y_2^8y_3^2 }{ y_1^{18}}     
-{\frac {24057537075}{32768}}                       \,\frac{ y_2y_3^2y_8 }{ y_1^{12} }     \\ \T
+{\frac {7044102065}{32768}}                        \,\frac{ y_2y_4y_9 }{ y_1^{11}}     
-{\frac {20008618245}{16384}}                       \,\frac{ y_2y_3y_5y_6 }{ y_1^{12}  }     
+{\frac {74747945475}{16384}}                       \,\frac{ y_2y_3^2y_4y_5 }{ y_1^{13}}     
-{\frac {42370573245}{32768}}                       \,\frac{ y_2y_3y_4y_7 }{ y_1^{12}  }    \\ \T 
-{\frac {69991007625}{131072}}                      \,\frac{ y_3^2y_4y_6 }{ y_1^{12}}     
+{\frac {5635204575}{32768}}                        \,\frac{ y_3y_4y_8 }{ y_1^{11}  }     
-{\frac {104068152405}{16384}}                      \,\frac{ y_2^9y_4 }{ y_1^{18}}     
+{\frac {861719557125}{32768}}                      \,\frac{ y_2^3y_3^3y_4 }{ y_1^{15}  }  \\ \T   
-{\frac {1909423425}{65536}}                        \,\frac{ y_5y_9 }{ y_1^{10}}     
+{\frac {25370560215}{32768}}                       \,\frac{ y_2^4y_9 }{ y_1^{13}  }     
+{\frac {5234922693}{32768}}                        \,\frac{ y_2^2y_{11} }{ y_1^{11}}     
-{\frac {5081656475}{131072}}                       \,\frac{ y_4^4 }{ y_1^{12}  }     \\ \T
-{\frac {3545717175}{262144}}                       \,\frac{ y_7^2 }{ y_1^{10}}     
+{\frac {6506875375}{786432}}                       \,\frac{ y_{13} }{ y_1^{9}  }     
+{\frac {4154848425}{180224}}                       \,\frac{ y_5^3 }{ y_1^{11}}     
-{\frac {8459871525}{32768}}                        \,\frac{ y_3^2y_5^2 }{ y_1^{12}  }   \\ \T  
+{\frac {4958959005}{65536}}                        \,\frac{ y_4^2y_7 }{ y_1^{11} }     
\end{eqnarray*}

\begin{figure}[h!] 
\hspace{1.1cm} 
\epsfig{figure=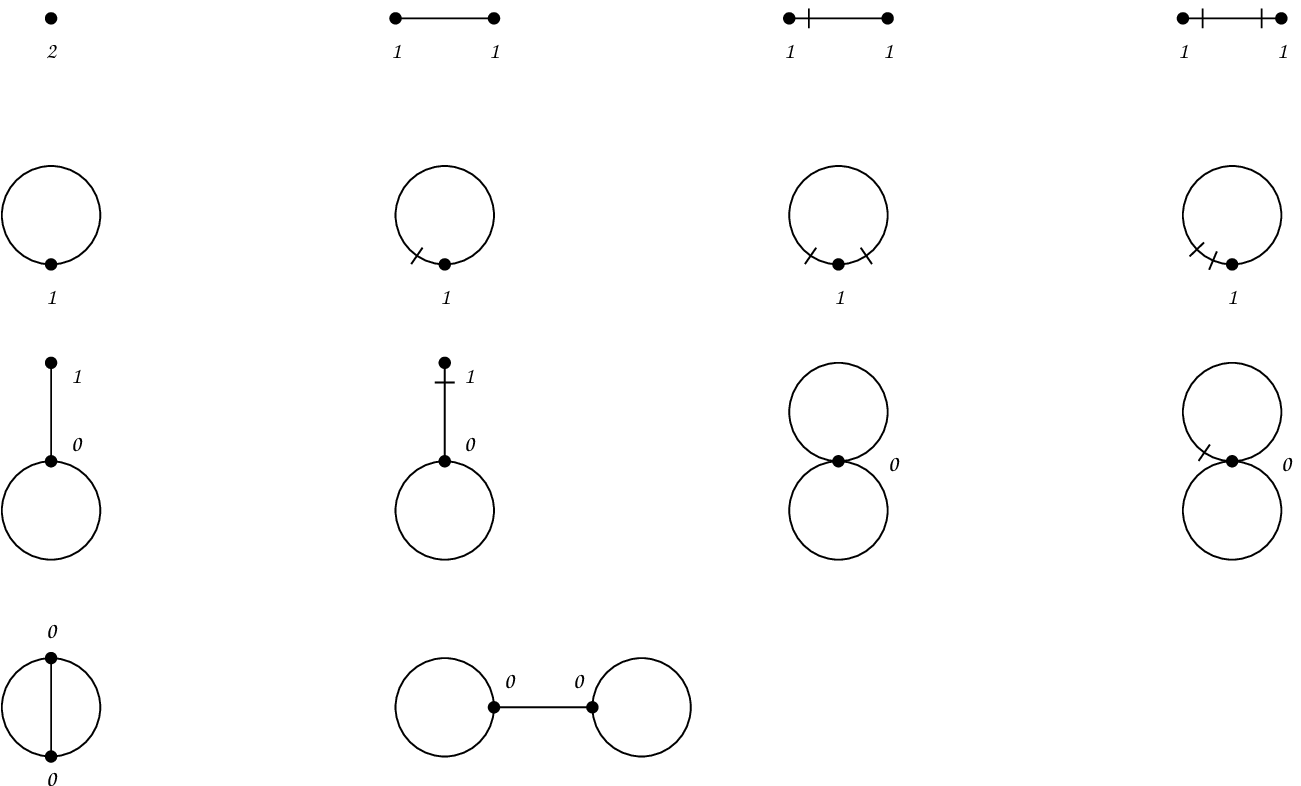} 
\caption{Free energy of order two. The number of small bars at each end of edge
  denotes the number of derivatives of the propagator. } 
\label{fig: Free energy of order two}
\end{figure}

\newpage
{\LARGE \bf Appendix B}\\ \\
$ \mathcal{F}^{(h)} $, the free energy at order $ h $, is equal to the
weighted sum of all
diagrams allowed by the effective Lagrange density $ \mathcal{L} $ with zero external legs,
where the sum of topological indices of the vertices (noted as small number
close to each vertex in figure \ref{fig: Free energy of order two}) plus the number of loops in the diagram is
equal to $ h $. Each diagram is weighted by its inverse symmetry factor (noted
as fraction in front of the diagrams in figure \ref{fig: Free energy of order two}). \\

\vspace{-10.22cm} 
\hspace{-1.7cm}
\begin{tabular}{lclclclcl}  
$ \D \mathcal{F}^{(2)} $  & $ = $ & \ \ \ \ \ \ \ \ \ \ \ \ \ \ \ \ \ \ \ \ \,  & + & $ \D
\frac{1}{2} $ & + & \ \ \ \ \ \ \ \ \ \ \ \ \ \ \ \ \ \ \ \ \,  & + & $
\D \frac{1}{2} $ \vspace{1.158cm} \\ 
& + & $ \D  \frac{1}{2} $  &  +   & \ \ \ \ \ \ \ \ \ \ \ \ \ \ \ \ \ \ \ \,
\  & + & $  \D  \frac{1}{2} $  &  + &   \ \ \ \ \ \ \ \ \ \ \ \ \ \ \ \ \ \,
\ \ \  \vspace{1.65cm} \\
& + & $  \D \frac{1}{2} $  & + & $ \D \frac{1}{2} $ & + & $ \D \frac{1}{8} $
& + &  $   \D \frac{1}{2} $ \vspace{1.64cm} \\
& + & $ \D \frac{1}{12} $  & + & $  \D \frac{1}{8} $  & & & &
\end{tabular}

\vspace{3cm}
The contribution to $ \mathcal{F}^{(2)} $ from the second diagram in the third
line of figure\nolinebreak\  \ref{fig: Free energy of order two}, for example, is given by:
\begin{displaymath}
\frac{1}{2} \sum_{i,j=1}^{2s} \lambda^{(1)}_{(0,1),i } \lambda^{(0)}_{ (3,0),j }
B_{i,j}^{1,0}  B_{j,j}^{0,0}.
\end{displaymath}

\newpage

\begin{figure}[h!] 
\hspace{1.1cm} 
\epsfig{figure=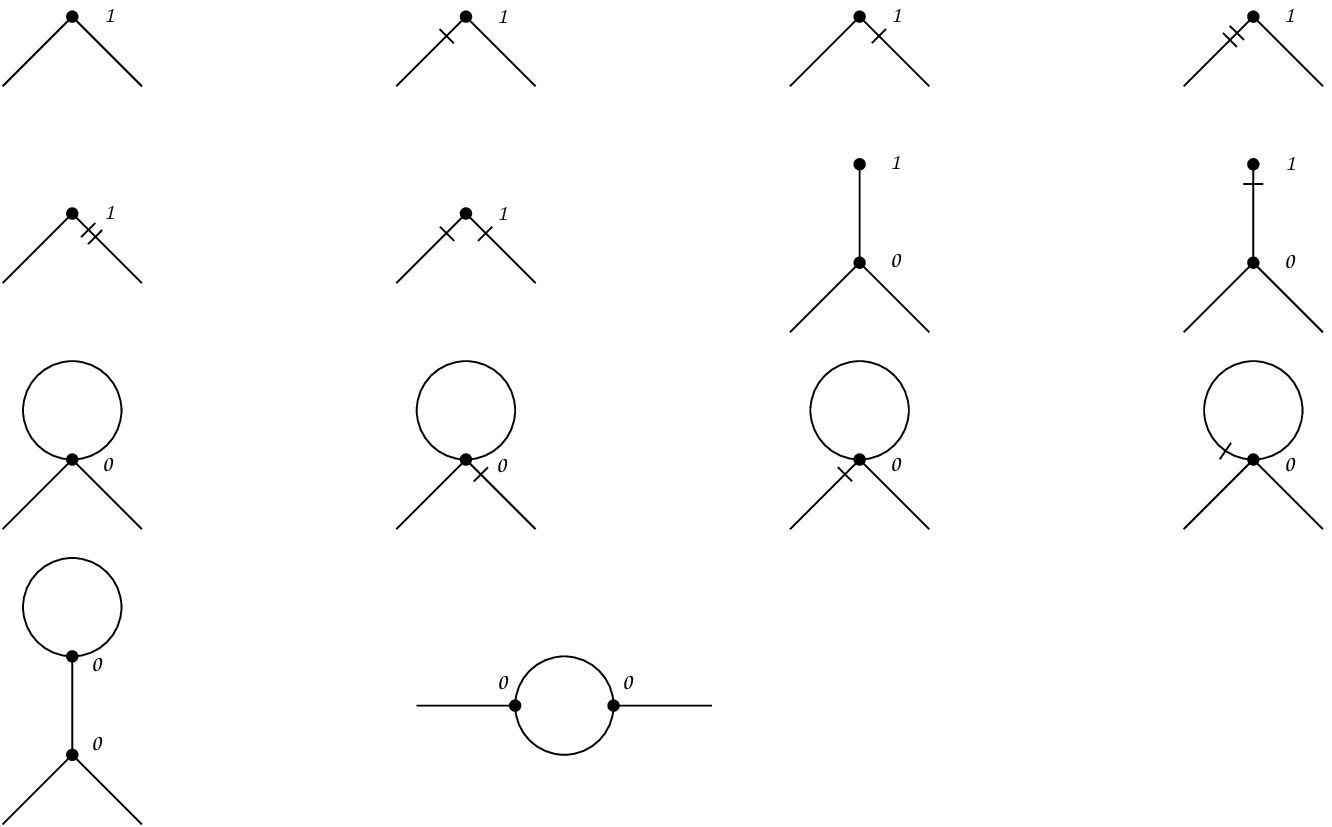} 
\caption{Correlation function $ W_2^{(1)} (p_1, p_2) $. The leg at the left
  side of each diagram belongs to $ p_1 $, the other to $ p_2 $.} 
\label{fig: W_2^1}
\end{figure}

\vspace{-10.42cm} 
\hspace{-1.7cm} 
\begin{tabular}{lclclclcl}  
$ \D W_2^{(1)} $  & $ = $ & \ \ \ \ \ \ \ \ \ \ \ \ \ \ \ \ \ \ \ \ \,  & + &
$ \D  $ & + & \ \ \ \ \ \ \ \ \ \ \ \ \ \ \ \ \ \ \ \ \,  & + & $
\D  $ \vspace{1.48cm} \\ 
& + & $ \D   $  &  +   & \ \ \ \ \ \ \ \ \ \ \ \ \ \ \ \ \ \ \ \,
\  & + & $  \D   $  &  + &   \ \ \ \ \ \ \ \ \ \ \ \ \ \ \ \ \ \,
\ \ \  \vspace{1.8cm} \\
& + & $  \D \frac{1}{2} $  & + & $ \D \frac{1}{2} $ & + & $ \D \frac{1}{2} $
& + &  $   $ \vspace{1.64cm} \\
& + & $ \D \frac{1}{2} $  & + & $  \D \frac{1}{2} $  & & & &
\end{tabular}

\vspace{3cm}
The contribution to $ W_2^{(1)} $ from the first diagram in the second
line of figure\nolinebreak\  \ref{fig: W_2^1}, for example, is given by:
\begin{displaymath}
\sum_{i=1}^{2s} \lambda^{(1)}_{(1,0,1),i } B_{i}^{0} (p_1)  B_{i}^{2} (p_2).
\end{displaymath} \\ \\

{\LARGE \bf Appendix C}\\ \\

This paragraph contains a formula for the number of terms in the Lagrange density $
\mathcal{L}_k^{(h)} $:
\begin{displaymath}
N(k,h):= | M_k^{(h)} | = \mbox{number of terms in } \mathcal{L}_k^{(h)} .
\end{displaymath}

 The number $ P(m,r) $ of multi-indices $ \alpha = (
\alpha_1, \ldots ,\alpha_m) \in ( \mathbb{N}_0 )^m $ fulfilling $ \sum_{j=1}^m
j \alpha_j = m $ and $ \sum_{j=1}^m \alpha_j = r $ can be computed \cite{Z-08} with

\begin{equation}
\sum_{m=0}^\infty \sum_{r=0}^\infty P(m,r) q^m x^r = \prod_{n=1}^\infty \left(
\frac{1}{1 - xq^n} \right) .
\label{eq: P(m,r)}
\end{equation}
Comparing to the conditions for $ M_k^{(h)} $ (section \ref{section:
  Lagrangean formalism}) one finds a summation
over $ m $ and a summation over $ r = k - \alpha_0 $:
\begin{displaymath}
N(k,h) =  \sum_{m=0}^{k+3h-3} \sum_{r=0}^k P(m,r) = \sum_{m=0}^{k+3h-3}
\sum_{r=0}^k \bigg( \frac{1}{m!} \frac{ \partial^m }{ \partial q^m } \bigg|_{q=0} \bigg) \left(
\frac{1}{r!} \frac{ \partial^r }{ \partial x^r } \bigg|_{x=0} \right) \prod_{n=1}^\infty
\left( \frac{ 1 }{ 1 - x q^n } \right) .
\end{displaymath}

\begin{figure}[h!] 
\begin{center}
\epsfig{figure=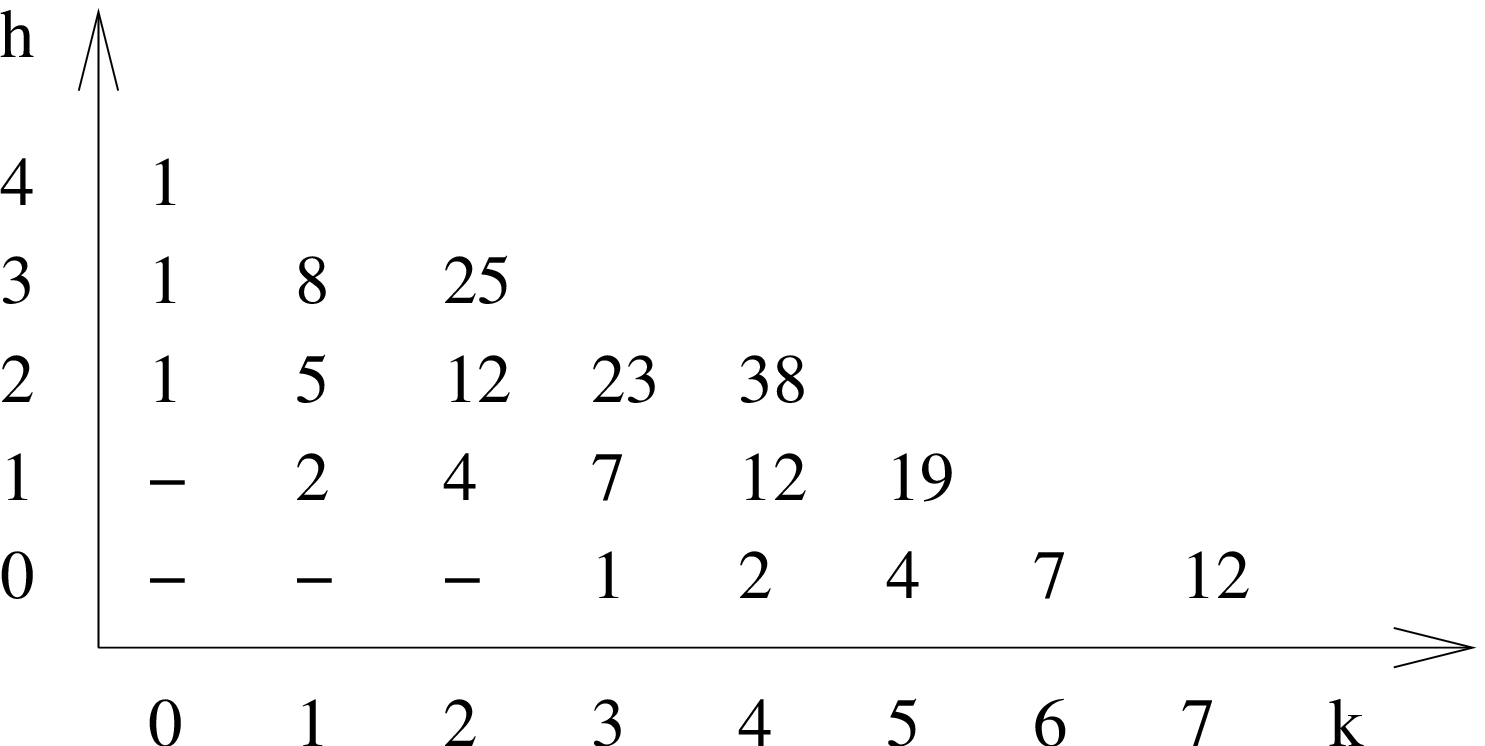, width=7cm} 
\caption{ Number of terms in the Lagrange density $ \mathcal{L}_k^{(h)} $ }
\end{center}
\end{figure}

\end{document}